\documentclass{ceurart}

\usepackage{booktabs}
\usepackage{multirow}
\usepackage{graphicx}
\usepackage{enumitem}
\usepackage{float}
\usepackage{url}
\usepackage{placeins}
\usepackage{subcaption}
\usepackage{tcolorbox}
\usepackage{array}
\usepackage{amsmath}
\usepackage{xcolor}
\usepackage{tabularx}

\begin{document}

\copyrightyear{2025}
\copyrightclause{
Copyright for this paper by its authors.
Use permitted under Creative Commons License Attribution 4.0
International (CC BY 4.0).
}

\conference{
IR-RAG @ SIGIR 2025: Workshop on Information Retrieval for Retrieval-Augmented Generation,
July 17, 2025, Padova, Italy
}

\title{Evaluating Modern RAG: Textual, Multimodal, Dense, and Late Interaction Pipelines}

\author[1]{Emre Kuru}[
orcid=0009-0007-6130-6272,
email=emre.kuru@ozu.edu.tr
]

\author[1]{Mehmet Onur Keskin}[
orcid=0000-0002-8025-4304,
email=onur.keskin@ozu.edu.tr
]

\address[1]{\"{O}zye\u{g}in University, Istanbul, T\"{u}rkiye}

\begin{abstract}
Retrieval-augmented generation (RAG) systems have traditionally relied on text-based pipelines that extract and retrieve information from documents. While efficient and lightweight, these approaches often struggle with documents where meaning is conveyed through layout, tables, and visual elements. Recent advances in multimodal pipelines, powered by vision-language models (VLMs), improve retrieval quality by jointly encoding visual and textual signals, but at increased computational and memory cost. We propose a quantitative, data-driven selection methodology that guides practitioners in choosing the most appropriate RAG pipeline for a given document corpus based on empirical effectiveness and resource constraints. We evaluate contemporary textual and multimodal pipelines, including dense and late-interaction architectures, analyze their trade-offs, and provide actionable guidance for balancing retrieval performance with system efficiency.
\end{abstract}


\maketitle

 \section{Introduction}
\label{sec:intro}

Retrieval-Augmented Generation (RAG) systems have emerged as a powerful solution for grounding large language models in external document corpora, particularly in knowledge-intensive domains such as finance, healthcare, and law \cite{setty2024improving,ng2025rag,pipitone2024legalbench}. Early RAG pipelines operated on purely textual representations in such domains, extracting content from scanned documents using Optical Character Recognition (OCR)~\cite{smith2007overview}. These \textit{text-only pipelines} proved effective for linear documents with clean structure, offering fast and lightweight retrieval. Although widely adopted due to their simplicity and efficiency, traditional text-only pipelines face significant limitations when applied to visually complex documents. OCR tends to flatten structural and spatial information, such as table layouts, section headers, or figure references, into unstructured text, often leading to the loss of semantic cues critical for accurate retrieval~\cite{zhang2024ocr, yimyam2020enhancing}. As a result, such pipelines frequently underperform tasks that require an understanding of document layout, tabular reasoning, or visual context.

Several enhanced text-based methods have been proposed to mitigate these limitations, such as layout-aware approaches. That incorporates structural information by applying document layout parsers \cite{wu2019detectron2, doclaynet2022} to segment documents into semantically meaningful regions such as titles, captions, tables, and figures, and using these annotations to guide chunking strategies \cite{yepes2024financial}, providing a partial structural understanding of the document. Furthermore, image captioning techniques have been used to convert visual elements (e.g., charts or figures) that cannot be captured by OCR into descriptive text~\cite{zhao2023retrieving}, allowing them to be indexed and retrieved within a traditional text-only framework. These enhancements extended the reach of text-based pipelines into visually rich domains while maintaining relatively low computational overhead. Nonetheless, even with such augmentations, text-based approaches still operate on symbolic approximations of visual content. They remain constrained by the quality of the layout or captioning models and often struggle to preserve fine-grained spatial relationships and formatting nuances. This led to the emergence of \textit{multimodal pipelines}~\cite{ma2024unifying, faysse2024colpali}, which operate directly on document images using vision-language models (VLMs). By encoding full-page inputs, these models jointly capture textual content, spatial layout, and visual features in a single embedding, dramatically improving retrieval in visually complex settings. However,  they come with steep system-level costs, including increased memory consumption, indexing, and retrieval latency.

In parallel to this modality shift, retrieval architectures have also evolved. Traditional retrieval architectures have embedded these documents (text or image) into a single vector, enabling fast similarity search. However, such representations compress the information, leading to diminished retrieval precision, especially when fine-grained semantic or structural alignment is required. To address this, \textit{late interaction} architectures~\cite{khattab2020colbert} have been proposed. Instead of encoding the input into a single vector, these models preserve multiple sub-representations and perform more expressive similarity computations at retrieval time. While this boosts performance, particularly in layout-sensitive queries, it also introduces higher latency, increased memory usage, and larger index sizes.

Despite the growing variety of pipeline designs, there remains little guidance on selecting the best approach for a specific task. This work addresses this gap by comprehensively evaluating state-of-the-art RAG retrieval pipelines across the two key design axes discussed: modality (text vs. vision) and architecture (dense vs. late interaction). Our study benchmarks these pipelines on retrieval performance and across a range of Quality-of-Service (QoS) metrics on two challenging, real-world datasets that exhibit diverse document structures and layout complexity. Building on these findings, we propose a selection methodology for identifying the most suitable RAG pipeline for a given document corpus and computational constraints. This is a deployment-oriented framework for selecting the most efficient and accurate pipeline.  The rest of the paper is organized as follows. Section~\ref{sec:related} reviews the related work. Section~\ref{sec:proposed} describes the retrieval pipelines evaluated in this study. Section~\ref{sec:eval} presents the experimental setup and results across effectiveness, latency, and memory usage. Finally, Section~\ref{sec:conclusion} concludes the paper with key takeaways and future directions.

\section{Related Work}
\label{sec:related}

Retrieval-augmented generation (RAG) has become a central paradigm for enhancing large language models with access to external knowledge sources. Early RAG pipelines operated primarily over textual inputs derived via Optical Character Recognition (OCR), with minimal attention paid to document layout or visual structure. Lin and Byrne introduced one of the earliest RAG architectures based on OCR outputs, integrating Dense Passage Retrieval (DPR) with a generative language model for open-domain question answering over scanned texts~\cite{lin2022retrieval}. Their work demonstrated that standard dense retrieval methods could be extended to semi-structured documents with text extraction. To better handle structurally rich documents, several works have incorporated layout-aware techniques. Yepes \textit{et al.} proposed a structure-guided chunking strategy that segments documents into semantically distinct regions, such as titles, body text, and tables, using computer vision and NLP methods, thereby preserving some spatial information during retrieval~\cite{yepes2024financial}. However, OCR-based pipelines, even with such enhancements, remain constrained by transcription errors and the inherent flattening of spatial layout. Zhang \textit{et al.} showed that such limitations can significantly degrade both retrieval accuracy and generation quality, especially in documents where meaning is intrinsically tied to structure~\cite{zhang2024ocr}.

Recognizing these persistent limitations, recent work has explored multimodal retrieval pipelines that directly encode visual signals. Chen \textit{et al.} introduced MuRAG, a memory-augmented architecture capable of reasoning over both image and text embeddings~\cite{chen2022murag}. Ma \textit{et al.} proposed a unified embedding space for document screenshots, enabling dense retrieval over raw visual input~\cite{ma2024unifying}. Similarly, Riedler \textit{et al.} applied CLIP-based retrieval to visually complex documents, demonstrating the benefits of joint vision-language representations~\cite{riedler2024beyond}. Furthermore, Faysse \textit{et al.} introduced ColPali, a visual late interaction architecture that extends the ColBERT paradigm to full-page images~\cite{faysse2024colpali}. By retaining patch-level representations and applying token-wise similarity scoring at query time, ColPali achieves state-of-the-art retrieval performance on layout-sensitive benchmarks, exemplifying the performance gains achievable with sophisticated multimodal, late-interaction approaches.

While this body of work has significantly advanced retrieval performance across various modalities, most evaluations have focused narrowly on effectiveness metrics such as nDCG and Recall. As RAG systems become more powerful and complex, understanding their system-level implications—such as indexing latency, retrieval latency, and memory footprint—becomes paramount for practical deployment, especially at scale. In contrast to prior studies, our work offers a holistic evaluation of retrieval pipelines. We systematically compare dense and late interaction methods across both text and visual modalities, not only in terms of retrieval effectiveness but also with respect to these crucial Quality-of-Service (QoS) metrics, which are essential for real-world deployment scenarios. To our knowledge, this is one of the first comprehensive analyses directly comparing these four pipeline archetypes—Text-Dense, Text-Late Interaction, Visual-Dense, and Visual-Late Interaction—across both retrieval quality and detailed operational costs.

\section{Proposed Approach}
\label{sec:proposed}

\begin{figure}[h]
    \centering
    \includegraphics[width=0.65\linewidth]{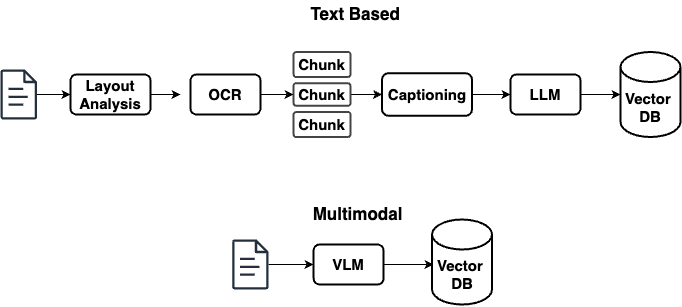}
    \caption{Indexing Workflows of Text \& Multimodal Pipelines}
    \label{fig:indexing-workflows}
\end{figure}

This work proposes a data-driven selection methodology for identifying the most suitable RAG retrieval pipeline given a document corpus and its computational constraints. This approach is grounded in a comprehensive evaluation of the state-of-the-art pipelines that represent key axes in modern retrieval system design: input modality (textual vs. visual) and retrieval architecture (dense vs. late interaction). This section outlines the design decisions behind these pipelines, detailing how they preprocess and embed documents and how their underlying architectures handle retrieval. Figure~\ref{fig:indexing-workflows} illustrates the differences between text-based and multimodal pipelines. Text-based pipelines typically follow a multi-stage preprocessing procedure. Raw document pages are first passed through Optical Character Recognition (OCR) to extract textual content. Afterward, the layout analysis step is applied, segmenting the document into semantically meaningful regions such as tables, titles, and figures. These regions inform layout-aware chunking strategies that divide the extracted text into retrieval-ready text chunks. An image captioning module is applied to generate natural language descriptions that can be indexed alongside the text for chunks containing non-textual content, such as images or tables. Finally, each chunk is embedded using a language model and stored in a vector database.

In contrast, multimodal pipelines operate directly on full-page document images. These systems leverage a vision-language model (VLM) to process the entire page holistically, generating embeddings that capture textual, visual, and spatial cues in a unified representation. This end-to-end approach eliminates the need for OCR, chunking, or captioning, enabling a more direct and layout-aware document encoding. The resulting embeddings are stored directly in the vector database.

\begin{figure}[h]
    \centering
    \includegraphics[width=0.65\linewidth]{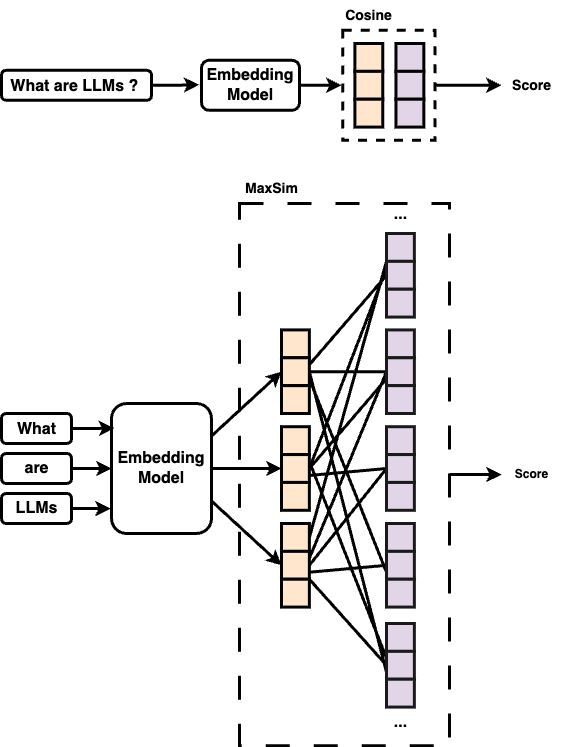}
    \caption{Dense vs Late Interaction Retrieval Architectures}
    \label{fig:retrieval-architecture}
\end{figure}

Figure~\ref{fig:retrieval-architecture} illustrates the architectural differences between dense and late interaction retrieval mechanisms. In dense retrieval, the document and query are encoded into single vector representations using a language or vision-language model. Retrieval is performed by computing the similarity between these vectors using a distance metric such as cosine similarity. This highly efficient approach scales well, making it well-suited for latency-sensitive applications. However, its compressed representation can lose fine-grained alignment, particularly important in structured or layout-dependent documents. Late interaction architectures address this limitation by preserving token-level or patch-level granularity during indexing. Instead of compressing the input into a single embedding, these models retain multiple embeddings per document, enabling more detailed and expressive similarity calculations at query time. A common scoring mechanism is MaxSim, where each query token is compared with all document tokens, and the maximum similarity scores are aggregated. While this approach improves retrieval precision, it also increases storage requirements and retrieval latency.

\subsection{Proposed Pipelines}

As discussed, we evaluate four retrieval pipelines that span the key design dimensions in modern RAG systems: input modality (text vs. vision) and retrieval architecture (dense vs. late interaction).

\vspace{4pt}
\noindent \textbf{Text-Dense.}
This pipeline implements a dense retrieval architecture of over-extracted text. As is common practice, we rely on the \texttt{Unstructured}\footnotemark[1] off-the-shelf tool in the highest resolution settings for OCR, layout analysis, and chunking (by-title). Afterward, for chunks that include non-textual elements, we set up a full-fledged captioning strategy, in which we feed the visual elements to a state-of-the-art Vision Language Model (Google-Gemini 2 0 Flash\footnotemark[2]) to obtain highly detailed textual descriptions of the elements. Each resulting chunk is then embedded into a dense vector using a text encoder.

\vspace{20pt}
\noindent  \textbf{Text-Late Interaction.}
This pipeline retains the same preprocessing strategy as the dense variant but uses a late interaction architecture to improve token-level alignment. Instead of embedding each chunk into a single vector, it preserves multiple sub-token embeddings, allowing fine-grained comparisons between query and document terms.

\vspace{4pt}
\noindent  \textbf{Visual-Dense.}
This pipeline skips OCR and processes full-page document images directly using a vision-language model. Each image is encoded into a dense vector that jointly captures visual and textual signals. This design allows for compact, layout-sensitive representations with minimal preprocessing.

\vspace{4pt}
\noindent  \textbf{Visual-Late Interaction.}
This pipeline operates on full-page images like the dense variant but uses a late interaction retrieval mechanism. It produces multi-vector embeddings for each document patch and performs expressive similarity scoring with patch-level granularity. This improves retrieval precision for layout-sensitive queries at the cost of greater computational overhead.

\footnotetext[1]{\url{www.unstructured.io}}

\footnotetext[2]{\url{https://ai.google.dev/gemini-api/docs/models}}

\section{Evaluation}
\label{sec:eval}

This section evaluates the proposed pipelines regarding retrieval performance across various QoS metrics, including memory consumption, indexing time, and retrieval latency. The remainder of this section is organized as follows. Section~\ref{subsec:setting} describes the experiment setting.Section~\ref{subsec:retrieval-performance} reports on retrieval effectiveness using standard ranking metrics to compare the pipelines' retrieval performances. Finally, Section~\ref{subsec:qos} presents our quality of service (QoS) evaluation, analyzing memory consumption, indexing, and retrieval latency.

\subsection{Expiriment Setting}
\label{subsec:setting}

\footnotetext[3]{\url{https://qdrant.tech}}

All experiments were conducted on a single NVIDIA H100 GPU with 80GB of memory. Each retrieval pipeline was evaluated independently using the same document corpus and query set to ensure a fair comparison. We implemented each pipeline under the same codebase and hardware to eliminate performance differences due to infrastructure. To ensure consistency across pipelines, all vector indices were stored using identical Qdrant\footnotemark[7] clusters.

\subsubsection*{Datasets.} 

We evaluate our retrieval pipelines using the REAL-MM-RAG benchmark~\cite{wasserman2025real}, a recent multimodal dataset designed for realistic document retrieval in structured domains. REAL-MM-RAG focuses on long, scanned documents with complex layouts. The queries are generated to reflect natural user questions rather than direct surface-level matches, providing a more rigorous test of retrieval models. To assess robustness under linguistic variation, each query in the dataset is rewritten up to three times using a large language model, resulting in four levels of query phrasing. These rephrased versions introduce increasing syntactic and lexical variation, allowing us to measure how well retrieval pipelines generalize beyond superficial token overlap and capture true semantic intent. We focus on two challenging subsets from this benchmark.

\vspace{2pt}
\textbf{FinReport} contains annual financial reports from IBM between 2005 and 2023. These documents include dense narrative content, structured financial tables, and visually segmented layouts. Queries typically require extracting specific values, understanding table structures, and reasoning over multiple modalities. This dataset is ideal for evaluating performance in layout-sensitive retrieval tasks.

\vspace{2pt}
\textbf{TechReport} consists of IBM technical documentation across systems like FlashSystem and Power Systems. These documents are primarily textual but include frequent visual components such as block diagrams, formatted tables. The queries in this domain target procedural details and technical descriptions, testing a model's ability to retrieve relevant content embedded in text and images.

\begin{figure}[h]
    \centering
    \begin{subfigure}[t]{0.6\linewidth}
        \centering
        \includegraphics[width=\linewidth]{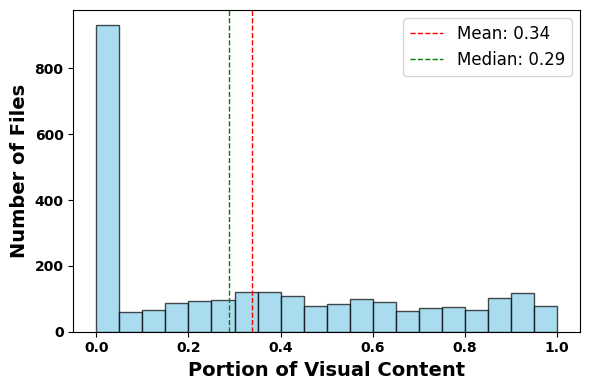}
        \caption{FinReport}
        \label{fig:visual_analysis_fin}
    \end{subfigure}
    
    \begin{subfigure}[t]{0.6\linewidth}
        \centering
        \includegraphics[width=\linewidth]{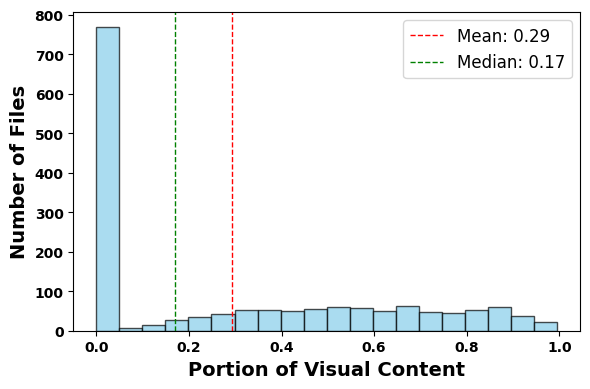}
        \caption{TechReport}
        \label{fig:visual_analysis_tech}
    \end{subfigure}
    
    \caption{Visual Component Rate Histogram across Datasets}
    \label{fig:visual_analysis_vertical}
\end{figure}

These datasets were chosen because they enable a fine-grained evaluation of how visual modality influences retrieval performance. Specifically, each document's proportion of visual content varies in both datasets, providing a balanced distribution of visually sparse and rich documents. This makes it possible to observe how retrieval pipelines perform under different visual conditions. As shown in Figure~\ref{fig:visual_analysis_vertical}, both datasets exhibit a broad range of visual content levels.

\subsubsection*{Models}

Each pipeline is built to represent the state-of-the-art, open-access models in their respective configurations. Specifically, text-based models were chosen from the top entries on the MTEB leaderboard \cite{enevoldsen2025mmteb}, while vision-based models were selected based on rankings in the ViDoRe leaderboard \cite{faysse2024colpali}. The following describes the specific models used for each pipeline and their underlying architectures:

\begin{itemize}

\item  \textbf{Baseline}: BM25~\cite{robertson2009probabilistic}, a sparse retrieval baseline widely used in information retrieval. BM25 scores documents based on exact token matches, using term frequency and inverse document frequency (TF-IDF) weighting. It does not use any learned parameters or embeddings.

\item \textbf{Text-single}: Linq-Embed-Mistral\footnotemark[4], is built upon the Mistral-7B architecture, fine-tuned for text retrieval tasks. It generates dense embeddings of 768 dimensions and supports a context window of 4096 tokens. The model size is approximately 7.11 billion parameters.

\item \textbf{Text-multi}: GTE-ModernColBERT-v1\footnotemark[5], a ColBERT-style embedding model optimized for fine-grained token matching. It produces 128-dimensional embeddings per token, resulting in multi-vector representations. The model comprises around 110 million parameters and supports a context window of 8192 tokens.

\footnotetext[4]{\url{https://huggingface.co/Linq-AI-Research/Linq-Embed-Mistral}}

\footnotetext[5]{\url{https://huggingface.co/lightonai/GTE-ModernColBERT-v1}}

\footnotetext[6]{\url{https://huggingface.co/nomic-ai/nomic-embed-multimodal-7b}}

\footnotetext[7]{\url{https://huggingface.co/Metric-AI/ColQwen2.5-7b-multilingual-v1.0}}

\item \textbf{Visual-single}: nomic-embed-multimodal-7b\footnotemark[6], a dense multimodal retriever based on the Bi-Qwen 2.5 architecture, an extension of the Qwen2.5-VL-7B vision-language model~\cite{yang2024qwen2}. It generates 768-dimensional dense embeddings per page. 

\item \textbf{Visual-multi}: ColQwen2.5-7b-multilingual-v1.0\footnotemark[7], a ColPali-style retriever built on the ColQwen 2.5 architecture, another extension of the Qwen2.5-VL-7B vision-language model~\cite{yang2024qwen2}. Unlike its dense counterpart, ColQwen produces multi-vector representations for text and images, enabling more fine-grained matching between queries and document regions. 

\end{itemize}

\subsection{Retrieval Performance}
\label{subsec:retrieval-performance}

Retrieval pipelines must ultimately be judged by their ability to return relevant content in response to a query. To quantify retrieval effectiveness, this section evaluates each pipeline using four widely adopted top-\(k\) ranking metrics (in our case, we chose \(k\) = 5): 

\begin{itemize}
    \item \textbf{nDCG@5 (Normalized Discounted Cumulative Gain):} Captures graded relevance by rewarding highly ranked relevant documents more than later in the result list.

    \item \textbf{MRR@5 (Mean Reciprocal Rank):} Reflects how early the first relevant document is retrieved, favoring pipelines that return relevant results near the top.
    
    \item \textbf{Precision@5:} Measures the proportion of retrieved documents within the top 5 that are relevant.

    \item \textbf{Recall@5:} Estimates the fraction of all relevant documents that appear in the top 5 results.
\end{itemize}

These metrics provide a comprehensive view of retrieval performance, balancing ranking position, accuracy, and completeness. Table~\ref{tab:retrieval_multiindex_levels} presents retrieval performance across four rephrasing levels for each pipeline and dataset. To assess the statistical significance of these results, we adopt a two-stage testing procedure. First, we apply group-level significance tests to determine whether any pipeline significantly differs from the rest. If the group contains three or more pipelines, we apply a \textit{Repeated Measures ANOVA} when normality is satisfied, and fall back to the \textit{Friedman test} otherwise. For pairwise comparisons, we apply either the \textit{Paired t-test} (for normally distributed and homogeneous samples) or the\textit{ Wilcoxon Signed-Rank Test} (for non-normal or heterogeneous samples). Normality is tested using the \textit{Kolmogorov-Smirnov test}, and homogeneity using \textit{Levene’s test}.

The \textit{visual-multi} pipeline significantly outperforms all other methods ($p < 0.01$) across all metrics and rephrasing levels, highlighting its robustness and superior generalization. Notably, even the \textit{visual-single} (dense vision-language) pipeline significantly outperforms all text-based pipelines across all rephrasing levels and metrics, confirming that multimodal pipelines; regardless of representation granularity, offer a clear and statistically significant advantage over purely textual models. Furthermore, while the baseline \textit{BM25} performs competitively on the original query set (occasionally surpassing some text-based pipelines) its performance degrades very sharply under rephrasing. This consistent drop underscores \textit{BM25’s} reliance on exact token matches and its vulnerability to lexical variation. Conversely, the proposed pipelines exhibit moderate performance drops as query rephrasing increases, reflecting the increased difficulty of matching semantically similar but lexically varied queries. However, this degradation is relatively consistent across architectures, with no single pipeline type disproportionately affected. These results suggest that vision-based retrieval maintains a consistent advantage even under paraphrasing stress, while all pipelines are sensitive to linguistic variation.

\begin{table}[t]
\centering
\caption{Retrieval Results Across Rephrase Levels Per Dataset}
\resizebox{0.95\textwidth}{!}{
\begin{tabular}{clcccccccc}
\toprule
\textbf{Level} & \textbf{Pipeline} & \multicolumn{4}{c}{\textbf{FinReport}} & \multicolumn{4}{c}{\textbf{TechReport}} \\
\cmidrule(r){3-6} \cmidrule(l){7-10}
& & nDCG@5 & MRR@5 & Precision@5 & Recall@5 & nDCG@5 & MRR@5 & Precision@5 & Recall@5 \\
\midrule
\multirow[c]{6}{*}{0} 
& BM25          & 0.43 $\pm$ 0.43 & 0.39 $\pm$ 0.43 & 0.11 $\pm$ 0.10 & 0.53 $\pm$ 0.50 & 0.61 $\pm$ 0.42 & 0.57 $\pm$ 0.43 & 0.14 $\pm$ 0.09 & 0.72 $\pm$ 0.45 \\
& text-single      & 0.46 $\pm$ 0.42 & 0.41 $\pm$ 0.42 & 0.12 $\pm$ 0.10 & 0.60 $\pm$ 0.49 & 0.61 $\pm$ 0.39 & 0.56 $\pm$ 0.41 & 0.16 $\pm$ 0.08 & 0.78 $\pm$ 0.42 \\
& text-multi       & 0.32 $\pm$ 0.43 & 0.30 $\pm$ 0.42 & 0.08 $\pm$ 0.10 & 0.40 $\pm$ 0.49 & 0.53 $\pm$ 0.45 & 0.50 $\pm$ 0.46 & 0.12 $\pm$ 0.10 & 0.62 $\pm$ 0.48 \\
& visual-single    & 0.66 $\pm$ 0.39 & 0.61 $\pm$ 0.41 & 0.16 $\pm$ 0.08 & 0.80 $\pm$ 0.40 & 0.73 $\pm$ 0.37 & 0.69 $\pm$ 0.39 & 0.17 $\pm$ 0.07 & 0.85 $\pm$ 0.36 \\
& visual-multi     & \textbf{0.74 $\pm$ 0.33} & \textbf{0.69 $\pm$ 0.37} & \textbf{0.18 $\pm$ 0.06} & \textbf{0.89 $\pm$ 0.31} & \textbf{0.85 $\pm$ 0.28} & \textbf{0.82 $\pm$ 0.32} & \textbf{0.19 $\pm$ 0.05} & \textbf{0.94 $\pm$ 0.24} \\
\midrule
\multirow[c]{6}{*}{1} 
& BM25     & 0.31 $\pm$ 0.40 & 0.28 $\pm$ 0.39 & 0.08 $\pm$ 0.10 & 0.41 $\pm$ 0.49 & 0.45 $\pm$ 0.44 & 0.42 $\pm$ 0.44 & 0.11 $\pm$ 0.10 & 0.56 $\pm$ 0.50 \\
& text-single      & 0.43 $\pm$ 0.42 & 0.39 $\pm$ 0.41 & 0.12 $\pm$ 0.10 & 0.58 $\pm$ 0.49 & 0.56 $\pm$ 0.40 & 0.50 $\pm$ 0.41 & 0.14 $\pm$ 0.09 & 0.72 $\pm$ 0.45 \\
& text-multi       & 0.28 $\pm$ 0.41 & 0.26 $\pm$ 0.41 & 0.07 $\pm$ 0.09 & 0.34 $\pm$ 0.47 & 0.48 $\pm$ 0.45 & 0.45 $\pm$ 0.45 & 0.12 $\pm$ 0.10 & 0.58 $\pm$ 0.49 \\
& visual-single    & 0.59 $\pm$ 0.42 & 0.54 $\pm$ 0.43 & 0.14 $\pm$ 0.09 & 0.72 $\pm$ 0.45 & 0.63 $\pm$ 0.40 & 0.59 $\pm$ 0.42 & 0.15 $\pm$ 0.08 & 0.77 $\pm$ 0.42 \\
& visual-multi     & \textbf{0.68 $\pm$ 0.37} & \textbf{0.63 $\pm$ 0.39} & \textbf{0.17 $\pm$ 0.07} & \textbf{0.84 $\pm$ 0.37} & \textbf{0.78 $\pm$ 0.33} & \textbf{0.75 $\pm$ 0.37} & \textbf{0.18 $\pm$ 0.06} & \textbf{0.89 $\pm$ 0.31} \\
\midrule
\multirow[c]{6}{*}{2} 
& BM25          & 0.22 $\pm$ 0.37 & 0.20 $\pm$ 0.36 & 0.06 $\pm$ 0.09 & 0.29 $\pm$ 0.45 & 0.38 $\pm$ 0.42 & 0.34 $\pm$ 0.41 & 0.10 $\pm$ 0.10 & 0.49 $\pm$ 0.50 \\
& text-single      & 0.40 $\pm$ 0.41 & 0.35 $\pm$ 0.41 & 0.11 $\pm$ 0.10 & 0.54 $\pm$ 0.50 & 0.53 $\pm$ 0.41 & 0.48 $\pm$ 0.42 & 0.14 $\pm$ 0.09 & 0.69 $\pm$ 0.46 \\
& text-multi       & 0.27 $\pm$ 0.40 & 0.24 $\pm$ 0.39 & 0.07 $\pm$ 0.09 & 0.34 $\pm$ 0.47 & 0.46 $\pm$ 0.44 & 0.42 $\pm$ 0.44 & 0.11 $\pm$ 0.10 & 0.56 $\pm$ 0.50 \\
& visual-single    & 0.52 $\pm$ 0.42 & 0.47 $\pm$ 0.43 & 0.13 $\pm$ 0.09 & 0.66 $\pm$ 0.47 & 0.58 $\pm$ 0.42 & 0.53 $\pm$ 0.43 & 0.14 $\pm$ 0.09 & 0.71 $\pm$ 0.45 \\
& visual-multi     & \textbf{0.61 $\pm$ 0.40} & \textbf{0.55 $\pm$ 0.42} & \textbf{0.15 $\pm$ 0.09} & \textbf{0.76 $\pm$ 0.43} & \textbf{0.75 $\pm$ 0.35} & \textbf{0.72 $\pm$ 0.38} & \textbf{0.17 $\pm$ 0.07} & \textbf{0.87 $\pm$ 0.34} \\
\midrule
\multirow[c]{6}{*}{3} 
& BM25          & 0.18 $\pm$ 0.34 & 0.16 $\pm$ 0.33 & 0.05 $\pm$ 0.09 & 0.24 $\pm$ 0.43 & 0.32 $\pm$ 0.41 & 0.29 $\pm$ 0.40 & 0.08 $\pm$ 0.10 & 0.40 $\pm$ 0.49 \\
& text-single      & 0.38 $\pm$ 0.41 & 0.34 $\pm$ 0.40 & 0.10 $\pm$ 0.10 & 0.51 $\pm$ 0.50 & 0.51 $\pm$ 0.41 & 0.46 $\pm$ 0.42 & 0.13 $\pm$ 0.09 & 0.67 $\pm$ 0.47 \\
& text-multi       & 0.23 $\pm$ 0.38 & 0.21 $\pm$ 0.37 & 0.06 $\pm$ 0.09 & 0.30 $\pm$ 0.46 & 0.42 $\pm$ 0.44 & 0.39 $\pm$ 0.43 & 0.11 $\pm$ 0.10 & 0.53 $\pm$ 0.50 \\
& visual-single    & 0.48 $\pm$ 0.43 & 0.43 $\pm$ 0.43 & 0.12 $\pm$ 0.10 & 0.61 $\pm$ 0.49 & 0.55 $\pm$ 0.42 & 0.50 $\pm$ 0.43 & 0.14 $\pm$ 0.09 & 0.68 $\pm$ 0.47 \\
& visual-multi     & \textbf{0.57 $\pm$ 0.41} & \textbf{0.52 $\pm$ 0.42} & \textbf{0.14 $\pm$ 0.09} & \textbf{0.72 $\pm$ 0.45} & \textbf{0.73 $\pm$ 0.36} & \textbf{0.68 $\pm$ 0.39} & \textbf{0.17 $\pm$ 0.07} & \textbf{0.85 $\pm$ 0.36} \\
\bottomrule
\end{tabular}}
\label{tab:retrieval_multiindex_levels}
\end{table}

All pipelines exhibit a moderate performance drop as query rephrasing increases, reflecting the increased difficulty of matching semantically similar but lexically varied queries. However, this degradation is relatively consistent across architectures, with no single pipeline type disproportionately affected. These results suggest that vision-based retrieval maintains a consistent advantage even under paraphrasing stress, while all pipelines are sensitive to linguistic variation. 

To illustrate the strengths and limitations of different retrieval pipelines, two qualitative case studies are examined. These examples highlight how performance is affected across modality (text vs. vision) and retrieval architecture (dense vs. late interaction). The first case study comes from our FinReport dataset and focuses on the impact of modality choice, particularly how OCR-induced errors can compromise retrieval accuracy in text-based pipelines. 

\begin{tcolorbox}[
  colback=white, colframe=black!20, boxrule=0.5pt,
  width=0.8\linewidth, center,
  title=\textbf{Case 1:} Correct Context, fonttitle=\bfseries
]

\textbf{Q}: What changed IBM's company debt from \textbf{2004 to 2005?}

\begin{tabularx}{\linewidth}{>{\raggedright\arraybackslash}X rr}
\toprule
Year & \textbf{2005} & \textbf{2004} \\
\midrule
Total company debt & \$22{,}641 & \$22{,}927 \\
Non-Global Financing debt & \$2{,}142 & \$607 \\
Non-Global Financing debt & & \\
\quad capitalization & 6.7\% & 2.1\% \\
\bottomrule
\end{tabularx}

\end{tcolorbox}

\begin{tcolorbox}[
  colback=white,
  colframe=black!20,
  boxrule=0.5pt,
  width=0.8\linewidth,
  center,
  title=\textbf{Case 1:} OCR output,
  fonttitle=\bfseries
]

\begin{tabularx}{\linewidth}{>{\raggedright\arraybackslash}X cc}
\toprule
Total company debt & \$22{,}641 & \$22{,}927 \\
Non-Global Financing debt & \$2{,}142 & \$607 \\
Non-Global Financing debt & & \\
capitalization & 6.7\% & 2.1\% \\
\bottomrule
\end{tabularx}

\end{tcolorbox}

In this example, the query asks about the change in IBM's total company debt between 2004 and 2005. However, the years in the corresponding financial table are entirely omitted during OCR preprocessing. As a result, both text pipelines fail to locate the correct context, retrieving irrelevant passages instead. In contrast, the visual pipelines operate directly on the document image and successfully place the correct page at the top of the ranking. This example underscores how early-stage OCR failures can cascade through text-based retrieval systems and highlights the robustness of vision-based models in preserving layout-dependent information.

The second case study examines the effects of retrieval architecture, particularly the distinction between dense and late interaction models in the face of linguistic variation. In this example, all pipelines successfully retrieve the correct document when the query explicitly includes the term bandwidth, which aligns closely with the vocabulary used in the document. However, when the query is rephrased using the semantically equivalent phrase ``maximum amount of data that can be transferred per unit of time'', only the late interaction models retain the correct context at the top rank. The dense models, in contrast, fail to match the rephrased version due to their reliance on surface-level term overlap in fixed vector representations. This case highlights a key limitation of dense retrieval in real-world scenarios where query phrasing is highly variable. It demonstrates how late interaction architectures offer greater robustness by enabling finer-grained token-level comparisons that better capture semantic nuance.

\begin{tcolorbox}[colback=white, colframe=black!20, boxrule=0.5pt, title= \textbf{Case 2:} Relevant Document Context, fonttitle=\bfseries]
With this setting, the Bronze customers can reach up to 1,000 Mbps, but they have to share the maximum \textbf{bandwidth} with all the other customers at the Bronze level. The Silver customers can reach up to 1,000 Mbps and do not have to share their limit with other customers in their performance class. The Gold customers are unlimited because they are not part of any performance class.
\end{tcolorbox}

\begin{tcolorbox}[colback=white, colframe=black!20, boxrule=0.5pt, title=\textbf{Case 2:} Original Query, fonttitle=\bfseries]
What is the \textcolor{green!50!black}{\textbf{bandwidth}} limit for Bronze customers in IBM FlashSystem A9000?
\end{tcolorbox}

\begin{tcolorbox}[colback=white, colframe=black!20, boxrule=0.5pt, title=\textbf{Case 2:} Rephrased Query, fonttitle=\bfseries]
In the IBM FlashSystem A9000, what is the \textcolor{red!50!black}{\textbf{maximum amount of data that can be transferred per unit of time}} for users with a Bronze subscription?
\end{tcolorbox}

\begin{figure}[h]
    \centering
    \begin{subfigure}[h]{\linewidth}
        \centering
        \includegraphics[width=0.87\linewidth]{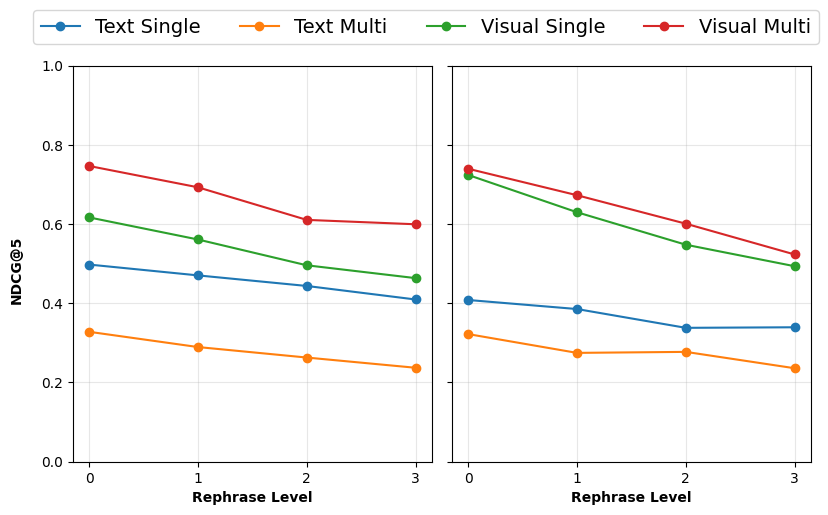}
        \caption{FinReport}
        \label{fig:text_vs_visual_fin}
    \end{subfigure}

    \begin{subfigure}[h]{\linewidth}
        \centering
        \includegraphics[width=0.87\linewidth]{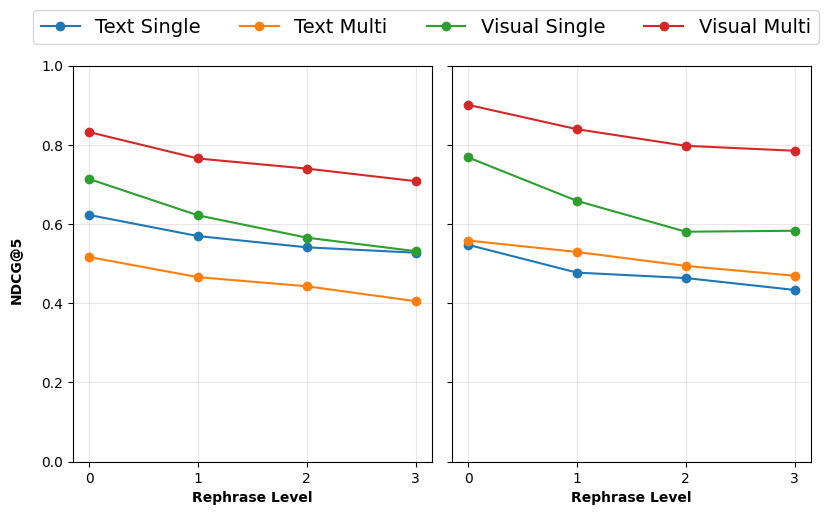}
        \caption{TechReport}
        \label{fig:text_vs_visual_tech}
    \end{subfigure}
    
    \caption{Retrieval Performance across Rephrase Levels for Text-Only (left) and Visual-Only (right) Queries for both Datasets}
    \label{fig:text_vs_visual_rephrasing}
\end{figure}

Figure~\ref{fig:text_vs_visual_rephrasing} breaks down retrieval performance by answer modality, referring to the type of document component in which the relevant information was embedded. Queries are labeled as \textit{text-only} when the answer resides within textual elements such as paragraphs or titles, and as \textit{visual-only} when the answer is found within visual elements like tables or images.  This figure further dissects the pipeline strengths. For "text-only" queries (left column of subplots for each dataset), we anticipate that while all pipelines might perform reasonably well, the \textbf{Visual-multi} and \textbf{Visual-single} pipelines leveraging advanced VLMs may still hold an edge due to their holistic page understanding, potentially capturing contextual cues even from surrounding visual layout that text-only models might miss. The \textbf{Text-single} and \textbf{Text-multi} pipelines are expected to be strong contenders here, with \texttt{Text-multi} potentially showing greater robustness to rephrasing due to its late-interaction mechanism. Conversely, for "visual-only" queries (right column of subplots), a more significant performance gap is expected. \textbf{Visual-multi} should dominate, adeptly handling queries requiring interpretation of tables or diagrams directly. \textbf{Visual-single} is also expected to perform well, significantly outperforming text-based pipelines. The text-based pipelines (\textbf{Text-single} and \textbf{Text-multi}) will likely struggle considerably on "visual-only" queries, their success is highly dependent on the quality of OCR and any generated captions for those visual elements. Performance degradation across rephrasing levels (0 to 3) is anticipated for all query types and pipelines, but the visual pipelines, particularly \textbf{Visual-multi}, might exhibit more graceful degradation on "visual-only" queries due to their richer representations. Differences between FinReport and TechReport would also be telling: the more table-intensive FinReport dataset is likely to highlight the superiority of visual pipelines for "visual-only" queries, whereas in TechReport, with its mix of text and diagrams, the distinctions might be nuanced but still favor visual approaches for visual queries.

\FloatBarrier

Figure~\ref{fig:visual_analysis_vertical} breaks down retrieval performance (nDCG@5) by document visual content proportion, with subplots for rephrasing levels (0 to 3). Visual-multi leads, especially in TechReport. Text-single is comparable to Visual-single on mostly textual documents but degrades sharply with more visual content (especially in FinReport). Interestingly, Text-multi, despite underperforming Visual-single in low-visual contexts, shows some resilience or even stronger results than Text-single in highly visual documents (FinReport, Level 0-1), suggesting its finer-grained token interactions might better navigate complex visual layouts indirectly via OCRed textual cues from those layouts.

\begin{figure}[h]
    \centering
    \begin{subfigure}[h]{\linewidth}
        \centering
        \includegraphics[width=\linewidth]{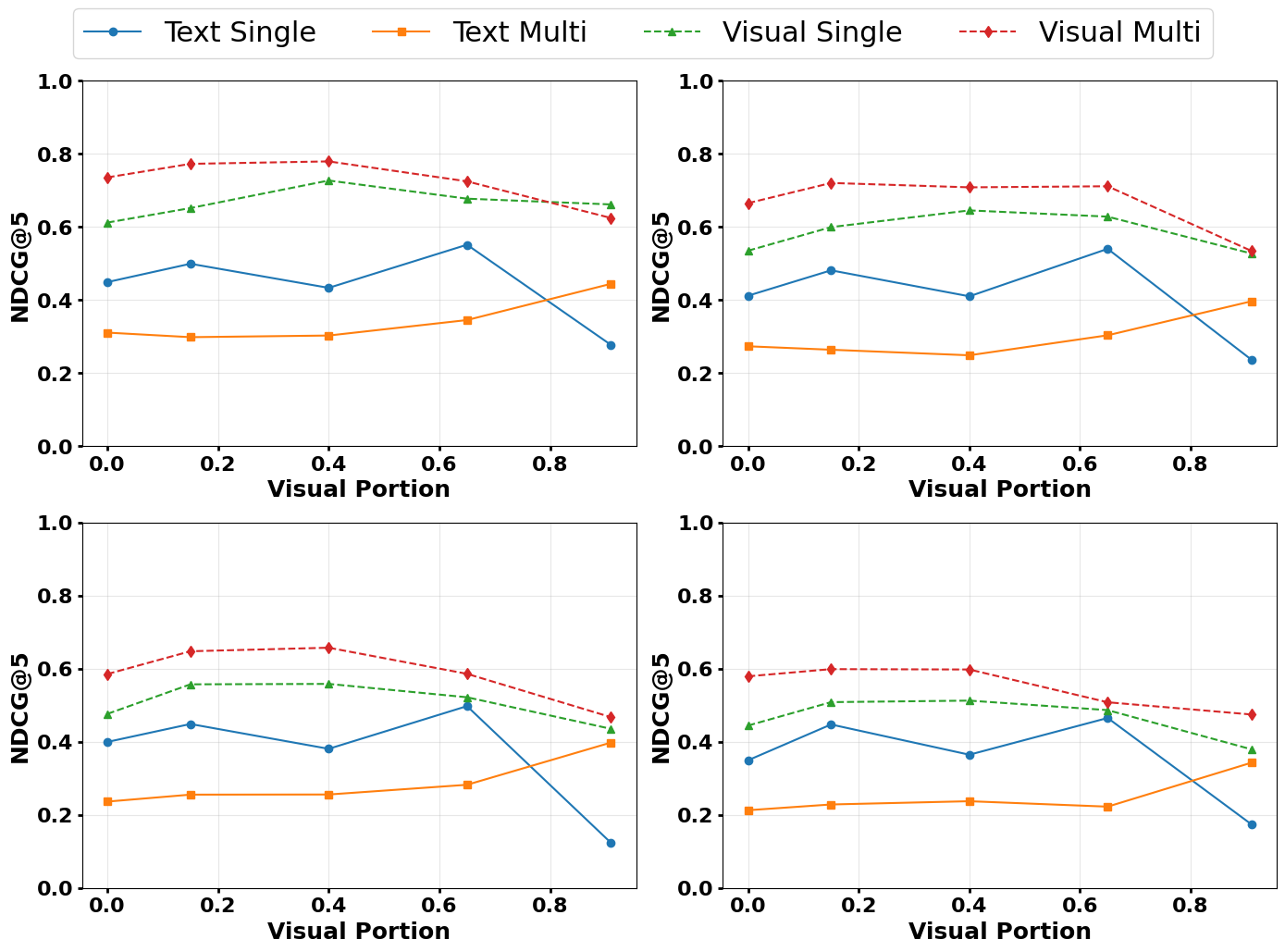}
        \caption{FinReport}
        \label{fig:visual_analysis_fin}
    \end{subfigure}

    \begin{subfigure}[h]{\linewidth}
        \centering
        \includegraphics[width=\linewidth]{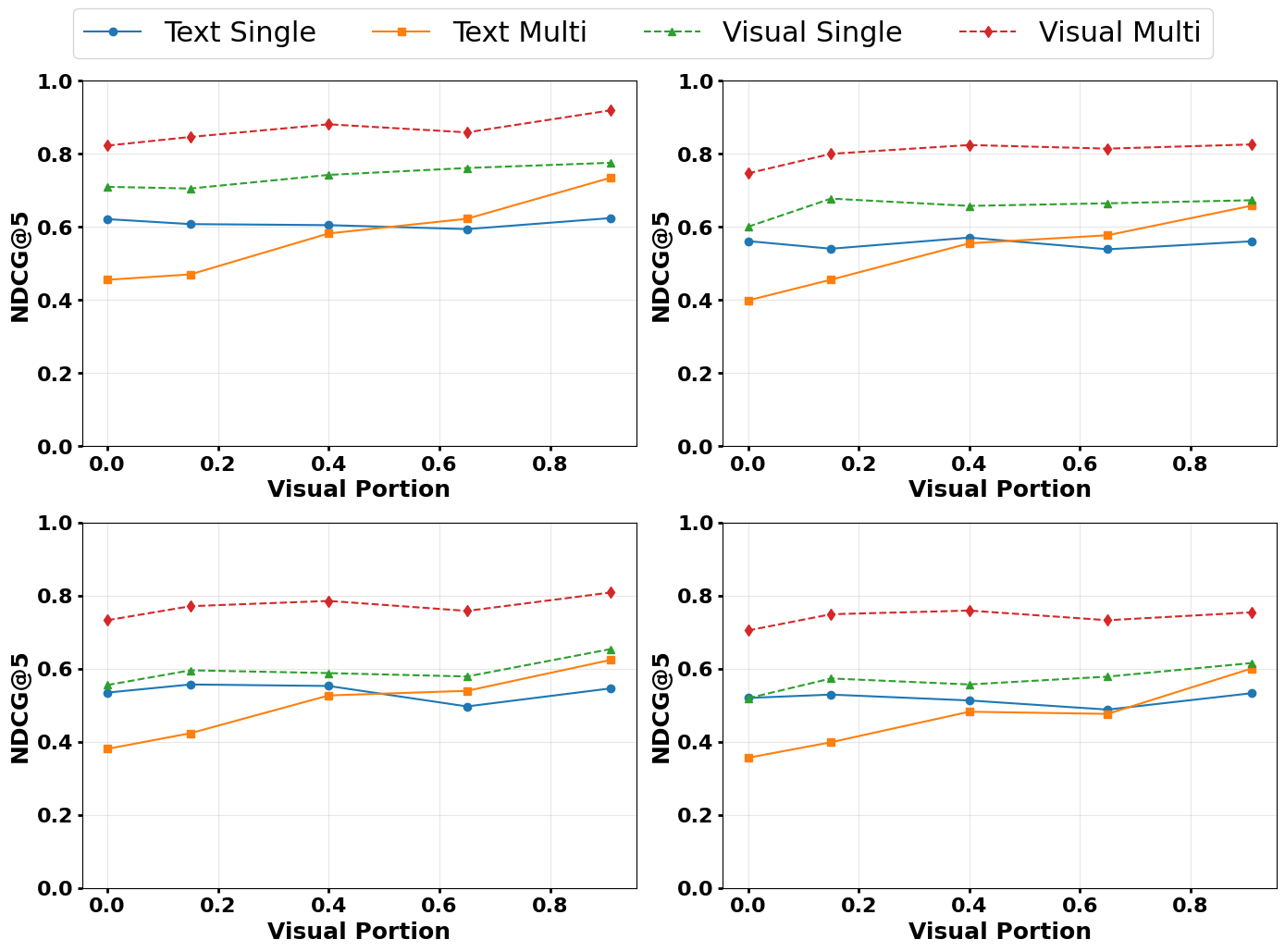}
        \caption{TechReport}
        \label{fig:visual_analysis_tech}
    \end{subfigure}
    
    \caption{Retrieval Performance across Visual Complexity for all Rephrase Levels (0-Top Left, 3-Bottom Right)}
    \label{fig:visual_analysis_vertical}
\end{figure}

\FloatBarrier

\subsection{Quality of Service (QoS)}
\label{subsec:qos}

Deployment of RAG systems demands careful consideration of system-level efficiency. To capture these concerns, each pipeline is evaluated across three key Quality of Service (QoS) dimensions: memory consumption, indexing latency, and retrieval latency. Memory consumption refers to the total size of the vector index stored in the database. Indexing latency measures the end-to-end time required to preprocess, embed, and upsert documents into the retrieval system. Retrieval latency captures the time needed to embed a query and retrieve the top-\(k\) most relevant documents. Together, these metrics reflect real-world constraints such as hardware cost, scalability, and responsiveness, factors critical when deploying retrieval systems over large document corpora.

Table~\ref{tab:memory_usage} presents the memory consumption of each pipeline. Due to their multi-vector representations, we can observe that late interaction models consume significantly more memory than their dense counterparts, which store embeddings at the token or patch level. This granularity allows for improved retrieval fidelity but comes at the cost of larger index sizes. Notably, \textit{visual-multi} exceeds 5GB in memory usage, more than 10 times that of the compact \textit{visual-dense} model. Interestingly, despite operating at higher embedding dimensionality, the visual-dense pipeline remains more storage-efficient overall. It encodes entire pages into a single vector, avoiding the cumulative overhead introduced by chunk-wise embeddings in text-based pipelines. Showcasing its scalability advantage in large-scale settings.

\begin{table}[h]
\centering
\caption{Total Memory Consumption for both Datasets}
\resizebox{0.4\linewidth}{!}{
\begin{tabular}{lc}
\hline
\textbf{Pipeline} & \textbf{Memory (MB)} \\
\hline
Text Single & 390.00 \\
Text Multi & 1079.00 \\
Visual Single & \textbf{340.00} \\
Visual Multi & 5068.80 \\
\hline
\end{tabular}
}
\label{tab:memory_usage}
\end{table}

Tables~\ref{tab:indexing_fin} and \ref{tab:indexing_tech} show the average indexing workflows for both our datasets, broken down into preprocessing, embedding, and upserting stages. As expected, text-based pipelines incur substantial preprocessing overhead due to OCR, layout parsing, and captioning. Despite this, their embedding and upsert times remain relatively lower, particularly for dense variants. In contrast, visual pipelines skip preprocessing entirely, enabling faster end-to-end indexing, especially for the dense variant. However, due to the higher dimensionality, the visual multi-pipeline requires significantly longer upsert durations. These results highlight a key trade-off: text pipelines are preprocessing-heavy but compact, while visual pipelines are plug-and-play but slower to upload.

\begin{table}[htb]
\centering
\caption{Average Indexing Workflow in Seconds - \textbf{FinReport}}
\resizebox{0.8\linewidth}{!}{
\begin{tabular}{lcccc}
\toprule
\textbf{Stage} & \textbf{Text Single} & \textbf{Text Multi} & \textbf{Visual Single} & \textbf{Visual Multi} \\
\midrule
Preprocess & 10.23 & 10.23 & \textbf{0.00}  & \textbf{0.00} \\
Embedding  & \textbf{0.02}  & 0.18  & 0.72  & 0.71 \\
Upsert     & \textbf{1.24}  & 3.26  & 5.69  & 6.04 \\
Total      & 11.49 & 13.67 & \textbf{6.41}  & 6.75 \\
\bottomrule
\end{tabular}
}
\label{tab:indexing_fin}
\end{table}

\begin{table}[htb]
\centering
\caption{Average Indexing Workflow in Seconds - \textbf{TechReport}}
\resizebox{0.8\linewidth}{!}{
\begin{tabular}{lcccc}
\toprule
\textbf{Stage} & \textbf{Text Single} & \textbf{Text Multi} & \textbf{Visual Single} & \textbf{Visual Multi} \\
\midrule
Preprocess & 10.70 & 10.70 & \textbf{0.00}  & \textbf{0.00} \\
Embedding  & 0.51  & \textbf{0.19}  & 0.91  & 1.07 \\
Upsert     & \textbf{1.89}  & 2.03  & 6.93  & 9.00 \\
Total      & 13.10 & 12.92 & \textbf{7.84}  & 10.07 \\
\bottomrule
\end{tabular}
}
\label{tab:indexing_tech}
\end{table}

Table~\ref{tab:retrieval_times} reports the average retrieval latency across all pipelines, broken down into query embedding and retrieval components. As expected, dense pipelines, \textit{text-single} and \textit{visual-single}, exhibit significantly lower latency, with total times under one second. In contrast, late interaction pipelines introduce a significant performance gap. While \textit{text-multi} remains within a negligible range, \textit{visual-multi} incurs exceptionally high retrieval times, exceeding 20 seconds per query. This discrepancy reflects the computational overhead of multi-vector retrieval at scale, particularly for high-dimensional vision-based embeddings.

\begin{table}[h]
\centering
\caption{Average Retrieval Latency in Seconds}
\resizebox{0.8\linewidth}{!}{
\begin{tabular}{lcccccc}
\toprule
\textbf{Pipeline} & \multicolumn{3}{c}{\textbf{TechReport}} & \multicolumn{3}{c}{\textbf{FinReport}} \\
\cmidrule(r){2-4} \cmidrule(l){5-7}
& Embedding & Retrieval & Total & Embedding & Retrieval & Total \\
\midrule
Text Single    & 0.04 & \textbf{0.50} & \textbf{0.54}  & 0.03 & \textbf{0.40} & \textbf{0.43} \\
Text Multi     & \textbf{0.01} & 2.99 & 3.01  & \textbf{0.01} & 4.19 & 4.20 \\
Visual Single  & 0.05 & 0.58 & 0.63  & 0.05 & 0.55 & 0.60 \\
Visual Multi   & 0.05 & 22.53 & 22.58 & 0.03 & 27.35 & 27.39 \\
\bottomrule
\end{tabular}
}
\label{tab:retrieval_times}
\end{table}

\section{Discussion}
\label{sec:conclusion}

Our comprehensive evaluation of textual and multimodal RAG retrieval pipelines across dense and late interaction architectures yields several notable trends and practical insights.

First, the Text-multi (late interaction) pipeline consistently underperformed not only its visual counterparts but also the simpler Text-dense pipeline across most retrieval effectiveness metrics, while still incurring higher memory usage and slower retrieval latency than Text-dense. For instance, it achieved a mean nDCG@5 of only 0.32 on the FinReport dataset (Level 0 rephrase), significantly lower than Text-single (0.46) and Visual-single (0.66).
In contrast, among the visual pipelines, Visual-multi (late interaction) achieved the highest retrieval performance overall, demonstrating the power of fine-grained, patch-level visual understanding. However, this superior effectiveness comes at a steep operational cost, particularly in terms of memory footprint (over 5GB) and retrieval latency (exceeding 20 seconds per query), making it potentially less suitable for resource-constrained or latency-sensitive scenarios.

Interestingly, our results challenge the common assumption that vision-based pipelines are invariably slower and more memory-intensive across the board. The Visual-single (dense) pipeline emerged as a compelling alternative, delivering significantly better retrieval performance than both text-based pipelines (e.g., 0.66 nDCG@5 on FinReport vs. 0.46 for Text-single and 0.32 for Text-multi) while also offering comparable or superior indexing throughput and the lowest memory consumption of all tested pipelines. This positions Visual-single as an excellent candidate for many real-world applications seeking a balance of effectiveness and efficiency.

Furthermore from our QoS results; we can observe that, embedding time itself was nearly negligible across all pipelines. The primary bottlenecks varied by modality: visual pipelines were predominantly constrained by the embedding and vector database upsert phase (especially Visual-multi), whereas text pipelines were bottlenecked by the extensive preprocessing phase (OCR, layout analysis, captioning).

\section{Future Work}

Our evaluation also revealed an important consideration for future benchmark development. Existing retrieval benchmarks, even those labeled as multimodal, often predominantly emphasize the retrieval of semantic answers that are textual, even if those answers are located within visual elements like charts or tables.
This can create a bias where the unique challenges and capabilities of retrieving genuinely non-textual, visually grounded information are underrepresented with text retrieval from images. Future work should aim to address this gap by developing more rigid evaluation methodologies and datasets that specifically target the retrieval of diverse visual information, moving beyond text embedded in images.

Complementing this direction, we also plan to explore a hybrid retrieval strategy that allocates documents to the most suitable indexing pipeline based on their structure and retrieval needs. For example, visually complex documents containing tables or figures may be better indexed using late-interaction visual models to improve retrieval performance, while simpler documents may be handled more efficiently with dense pipelines. At query time, the system retrieves results from all pipelines and fuses the outputs, aiming to reduce memory usage and retrieval latency without significantly compromising retrieval quality.

In summary, this study provides a data-driven foundation for selecting appropriate RAG retrieval pipelines. Practitioners can use these findings—weighing the trade-offs between visual/textual modality, dense/late-interaction architecture, retrieval effectiveness, and QoS metrics—to make informed decisions tailored to their specific document corpora, task requirements, and operational constraints. These insights contribute to the ongoing effort to build more efficient, effective, and practically deployable RAG systems.

\section*{Declaration on Generative AI}
The author(s) have not employed any Generative AI tools in the preparation of this work.

\bibliography{bibtex}

\end{document}